\documentclass[conference]{IEEEtran}

\usepackage{algorithmic}
\usepackage{graphicx}
\usepackage{textcomp}
\usepackage{xcolor}
\usepackage[hidelinks]{hyperref}
\usepackage{multirow}
\usepackage{booktabs}
\usepackage{nicefrac}
\usepackage{mdframed}
\mdfdefinestyle{summarybox}{%
  rightline=true,
  innerleftmargin=10,
  innerrightmargin=10,
  linewidth=2pt,
  topline=false,
  bottomline=false,
  skipabove=\topsep,
  skipbelow=\topsep,
  frametitlefont=\normalfont\sc
}
\def\BibTeX{{\rm B\kern-.05em{\sc i\kern-.025em b}\kern-.08em
    T\kern-.1667em\lower.7ex\hbox{E}\kern-.125emX}}
\newcommand{\gqb}{\textsc{GraphicalQueryBuilder}}
    
\begin{document}
\begin{sloppy}

\title{Querying Spatial-Temporal-Spectral Data Using a Graphical Query Builder}

\author{\IEEEauthorblockN{Adela Gorczynska}
\IEEEauthorblockA{\textit{The University of Adelaide}\\
Adelaide, SA \\
Australia}
\and
\IEEEauthorblockN{Peter Fule}
\IEEEauthorblockA{\textit{Swordfish Computing}\\
Adelaide, SA \\
Australia}
\and
\IEEEauthorblockN{Christoph Treude}
\IEEEauthorblockA{\textit{The University of Melbourne}\\
Melbourne, VIC \\
Australia}}

\maketitle

\begin{abstract}
Constructing complex queries on data which combines spatial, temporal, and spectral aspects is a challenging and error-prone process. Query interfaces of general-purpose database management systems fall short in providing intuitive support for users to effectively and efficiently construct queries. To address this situation, we developed \gqb, a tool which provides interactive and immediate feedback during query construction and visually represents search space and queries. A user study with ten professionals showed that users were able to complete queries on average 40\% faster and 40 percentage points more accurately compared to a Microsoft Access baseline.
\end{abstract}


\section{Introduction and Motivation}

Effectively constructing queries on advanced database systems that contain domain-specific data which does not follow basic numerical or textual patterns is challenging and error-prone~\cite{cavalcanti2006querying, reschenhofer2016supporting}. To support decision making in such scenarios, visual and interactive query systems have been proposed, often focusing on specific domains, such as spreadsheets~\cite{cunha2014sql}, sports~\cite{richly2018leveraging}, Linked Open Data~\cite{vega2016visual}, geographical data~\cite{dobesova2012comparison}, or spatio-temporal data~\cite{li2004interactive, emrich2012querying}. 

Many studies provide evidence that visual interfaces outperform textual interfaces in high-complexity scenarios. Speier et al.~\cite{speier2003influence} report that visual-based query interfaces have been shown to increase decision quality and decrease decision time as well as decrease confusion when entering a query. In addition, graphical displays are much easier to interpret and understand compared to a tabular display. Some examples of visual interfaces include presenting address data values on a map compared to a table format, using slider bars or buttons as opposed to entering manually, or providing an immediate visual display as users input their query. However, further research is needed on how to construct such visual and interactive query interfaces in many domains, such as for geospatial data~\cite{erskine2014business}.

In this paper, we introduce \gqb, a tool specifically designed for querying spatial-temporal-spectral data. To the best of our knowledge, domain-specific query tools for this kind of data do not currently exist, and professionals have to rely on general-purpose tools such as relational database queries in their decision making processes. Spatial-temporal-spectral data is common in domains such as remote sensing~\cite{zhang2018missing} and unmanned aerial vehicles (UAV)~\cite{chen2021geo}, with use cases such as finding space in the spectrum to put in a new radio tower, or identifying the source of interference on a radio,

We built \gqb~in collaboration with an industry partner working with a transmitter dataset. The dataset contains name, location, hours of operation, centre frequency, and bandwidth for the transmissions of a set of transmitters. \gqb~was designed to enable users to easily extract useful information including conflicts (e.g., interference between signals from different transmitters) and gaps within the data. The interface allows users to continually refine queries based on their needs and to view their queries as they are being built. This minimises the possibility of incorrect query input and clearly shows users what they are querying for. 

Our main contribution is the assessment of the effectiveness of the interplay of individual interface components in the spatial-temporal-spectral domain. We evaluated \gqb~in a user study with ten professionals which showed that they were able to complete a set of queries 40\% faster than with a Microsoft Access baseline and with much higher accuracy on their first attempt: \nicefrac{17}{20} queries were correctly entered on first try via \gqb, compared to \nicefrac{9}{20} with the Microsoft Access baseline.

\section{\gqb}

\begin{table*}
\caption{Sample data}
\label{tab:data}
\centering
\begin{tabular}{lr@{}r@{}r@{}r@{}r@{}r@{}rr@{}r@{}rrr}
\toprule
\multicolumn{1}{c}{Name} & \multicolumn{7}{c}{Location} & \multicolumn{3}{c}{Hours} & Centre Frequency & Bandwidth \\
\midrule
Mobile Phone Tower 123 & 38\textdegree & 40' & 11.86''~ & , & ~-90\textdegree & 7' & 9.73'' & 0:00 & -- & 24:00 & 900MHz & 3kHz \\
Railway Station Shortwave & 38\textdegree & 37' & 45.52''~ & , & ~-90\textdegree & 14' & 6.69'' & 5:00 & -- & 23:00 & 26MHz & 2kHz \\
Emergency Communications System & 38\textdegree & 37' & 36.84''~ & , & ~90\textdegree & 11' & 57.61'' & 0:00 & -- & 24:00 & 32Hz & 10kHz \\
Stadium & 38\textdegree & 37' & 58.90''~ & , & ~90\textdegree & 11' & 22.36'' & 8:00 & -- & 20:00 & 30MHz & 1kHz \\
International Aeronautical Distress &  &  &  &  &  &  &  & 00:00 & -- & 24:00 & 406.5MHz & 1kHz \\
University Satcom & 38\textdegree & 37' & 20.28''~ & , & ~90\textdegree & 13' & 57.76'' & 20:00 & -- & 10:00 & 2.564GHz & 15MHz \\
\bottomrule
\end{tabular}
\end{table*}

\begin{table*}
\caption{Sample tasks}
\label{tab:tasks}
\centering
\begin{tabular}{ll}
\toprule
\multirow{6}{*}{\rotatebox{90}{Simple}} & Anything that transmits at 90MHz +/- 1 MHz? \\
 & What frequencies are active between 1:00 and 4:00? \\
 & What frequencies are available between 3:00 and 8:00? \\
 & Anything that transmits within 1km of $\langle$Location$\rangle$? \\
 & What times are transmissions made in $\langle$Location$\rangle$? \\
 & Anything that does not transmit within the range 90 MHz--100 MHz? \\
\midrule
\multirow{6}{*}{\rotatebox{90}{Advanced}} & Anything that transmits in the range 78 MHz--85 MHz, within 1km of $\langle$Location1$\rangle$? \\
 & Anything that transmits within 1km of $\langle$Location1$\rangle$ or within 10km of $\langle$Location2$\rangle$? \\
 & Anything that transmits at 25 MHz, in the city of $\langle$Location1$\rangle$ at Noon? \\
 & Anything that transmits between hours 19:00--23:00 in the range 90 MHz--100 MHz or at 56000 +/- 8000 kHz? \\
 & Anything that transmits in the range 55MHz--60MHz, within 1km of $\langle$Location1$\rangle$ between 05:00--09:00? \\
 & Anything that transmits between hours 1:00--3:00, within 10km of $\langle$Location2$\rangle$ at 67500 kHz? \\
\bottomrule
\end{tabular}
\end{table*}

Table~\ref{tab:data} shows a small sample of the spatial-temporal-spectral data which \gqb~is optimised for. For each transmitter, the data contains a name (usually the name of the location of the transmitter), its exact location specified using longitude and latitude, its hours of operation, and the frequency range. The frequency range is either specified using the centre frequency and bandwidth (as for the examples in Table~\ref{tab:data}) or using the minmum and maximum frequency (not shown in the table). Note that the ranges for frequencies vary greatly from a few hertz to several gigahertz (i.e., $10^9$ Hz). Table~\ref{tab:tasks} shows example queries that need to be run on this data.

The following sections explain the design of \gqb, divided into front-end and back-end. Figure~\ref{fig:screenshot} shows a screenshot of the tool's user interface.

\subsection{Front-end}

\begin{figure*}
\centering
\includegraphics[width=\linewidth]{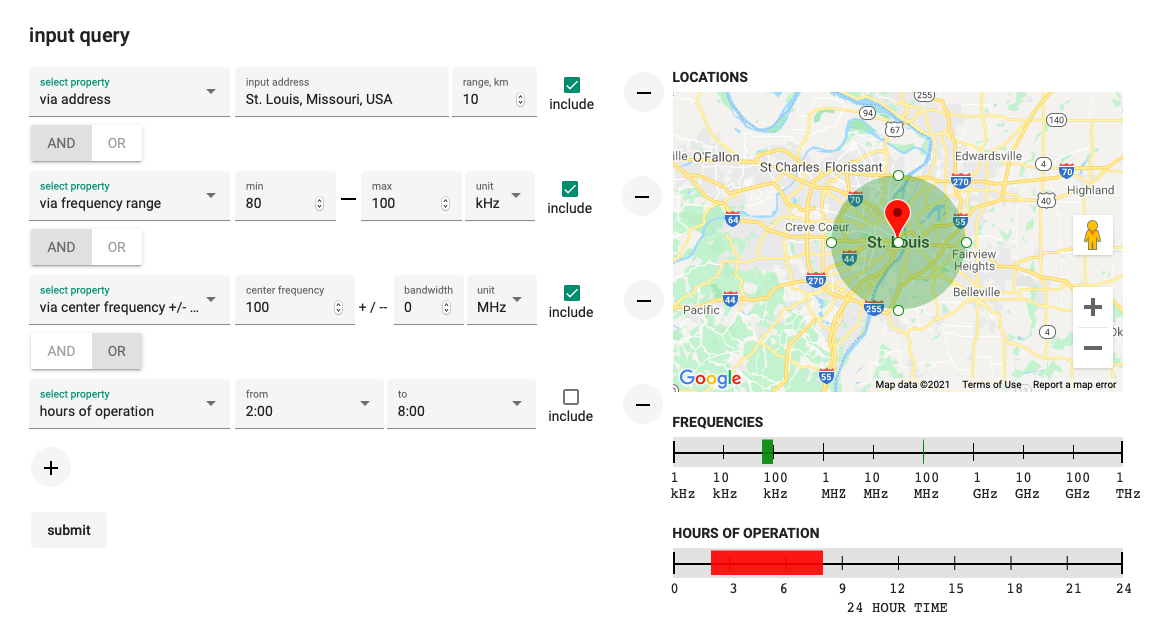}
\caption{User interface of \gqb}
\label{fig:screenshot}
\end{figure*}

The front-end was designed to satisfy the following requirements:

\begin{itemize}
\item Allow users to query by the transmitter's name, location, range from location, hours of operation, frequency and/or bandwidth.
\item Accommodate simple to advanced queries (i.e., search by two or more parameters).
\item Be intuitive to the extent that users can understand its behaviour without the need of special training or assistance.
\item Allow users to perform a query with minimum amount of effort.
\item Deliver expected results the first time rather than having users repeat the query (with variations) to get their intended result.
\end{itemize}

\gqb's front-end is implemented in Angular.\footnote{\url{https://angular.io/}} As shown in Figure~\ref{fig:screenshot}, the key feature of \gqb's user interface is the visualisation of the query as it is being built. This visual aspect allows users to verify that they are inputting the correct values of their query and it helps them understand what they are querying for.

The query component allows users to search by name, location, hours of operation, and frequency. Upon `name' selection, a drop-down field appears allowing users to choose the appropriate name. Users have the choice to enter `location' via address or latitude and longitude. A marker is placed on the map at the location entered. Users can enter their desired range in kilometres which appears as a circle on the respective marker. Users can also edit the circle on the map to adjust the range value. Upon `hours of operation' selection, a `from' and `to' hour field is displayed. Once users select from both fields, a visualisation representing the input hour range appears on the hours of operation axis. Finally, users have the choice to query by a `centre frequency' or a `frequency range'. A visualisation representing the input frequency appears on the frequency visualisation. Note the log scale to accommodate the wide range of frequencies from one hertz to one terahertz (i.e., $10^{12}$ Hz).

In addition, each query has an `include' checkbox, selected on default. If the query is included, the visualisation colour is green and if excluded, the visualisation colour is red. The query interface allows users to add properties to their query as they need, via the `+' button as well as remove properties, via the `--' button. Users can also select whether the query is joined by `AND' or `OR' by selecting either option.

\subsection{Back-end}

The back-end was designed to satisfy the following requirements:

\begin{itemize}
\item Efficiently store data in a relational database.
\item Retrieve and display expected data based on query
execution.
\end{itemize}

\gqb's backend is implemented in Django\footnote{\url{https://www.djangoproject.com/}} and connects to a relational PostgreSQL database.\footnote{\url{https://www.postgresql.org/}} The SQL statement is generated from the user interface web page and passed to the REST APIs defined in the Django application. Django will then send the appropriate response back to web page, built with Angular, in JSON format. The results are displayed in a table, see Figure~\ref{fig:results}.

\section{Evaluation Methodology}

To evaluate whether \gqb~is able to support users in querying spatial-temporal-spectral data, we asked two research questions:

\begin{description}
\item[\textbf{RQ1}] How \textit{accurately} can professionals formulate queries with \gqb~compared to a baseline?
\item[\textbf{RQ2}] How \textit{quickly} can professionals complete queries with \gqb~compared to a baseline?
\end{description}

To answer our research questions, we compared \gqb~against Microsoft Access,\footnote{\url{https://www.microsoft.com/en-au/microsoft-365/access}} a commercial database management system with a graphical user interface. Other specialised industry products, such as the Australian Communications and Media Authority (ACMA) search for registered transmitters,\footnote{\url{https://web.acma.gov.au/rrl/register_search.search_dispatcher}} did not contain details needed for our usage scenarios (e.g., time of operation). We recruited ten professionals from our industry partner to participate in a user study. All participants were familiar with SQL and the schema of the data. None of these professionals are authors of this paper or were involved in the design of \gqb. Each participant was asked to complete four tasks (see next paragraph), two with Microsoft Access and two with \gqb. 

Our industry partner provided us with several sample queries which they regularly use on spatial-temporal-spectral data. These queries were used as a guide while developing \gqb~and for the construction of a set of tasks for participants to complete during the evaluation. The tasks were designed so that they would test each property of \gqb~individually and well as combined with other properties. The tasks were similar in nature to the sample tasks shown in Table~\ref{tab:tasks}. Some tasks were made to be simple, by querying with a single condition, for the purpose of getting participants accustomed to both tools. An example simple query is ``Anything that transmits at 90MHz +/- 1 MH''. Other queries were more advanced, involving three conditions. An example advanced query is ``Anything that transmits between hours 1:00 -- 3:00, within 10km of $\langle$Location$\rangle$ at 67500 kHz''. Each participant completed one simple and one advanced task with each of the tools, for a total of four tasks per participant.

To evaluate whether the output of the users' query submissions was correct, the corresponding SQL statements were written and executed on the same data to compare the output. The time it took each user to complete each query was recorded and we asked participants about their overall experience after completing all queries. 

\section{Evaluation Results}

\begin{figure}
\centering
\includegraphics[width=\linewidth]{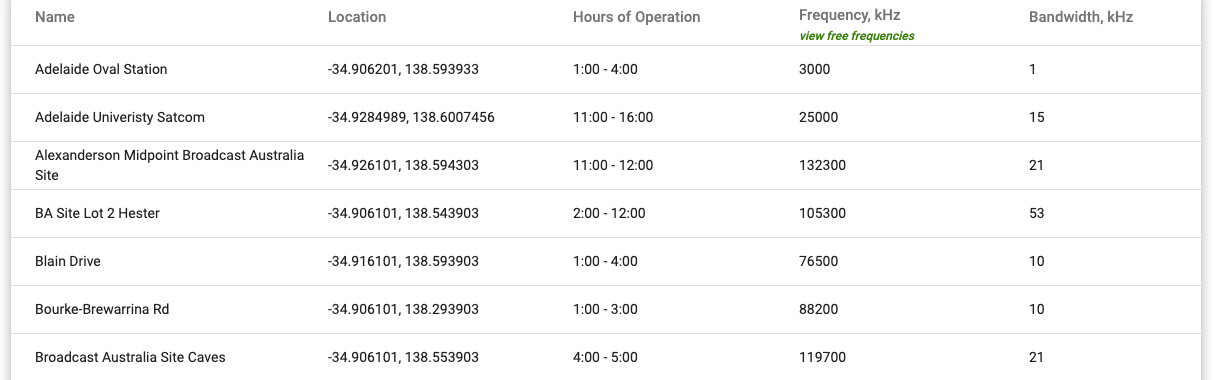}
\caption{\gqb~results display (excerpt)}
\label{fig:results}
\end{figure}

Participants were able to complete tasks on average 40\% faster with \gqb~compared to the Microsoft Access baseline (approximately two and a half minutes compared to approximately four minutes). In addition, only \nicefrac{9}{20} queries were correctly entered on first try via Microsoft Access compared to \nicefrac{17}{20} correctly entered queries via \gqb. Figure~\ref{fig:results} shows example output of \gqb.

All participants provided positive feedback with regard to \gqb's ease of use, intuitiveness, and efficiency. Some participants mentioned that current methods for data analysis are performed with tools similar to Microsoft Access.

During the evaluation, several suggestions were made for improvements to \gqb. One participant suggested giving users the ability to modify the input via visualisation elements other than the map. Another suggested including an option to export the results as a comma separated values (CSV) file. We will consider these suggestions for future versions of \gqb.

\begin{mdframed}[style=summarybox,frametitle=Summary]
In a user study to evaluate \gqb~with ten professionals completing four tasks each: 
\begin{itemize}
\item[\textbf{RQ1}] Users were able to enter queries 40 percentage points more accurately compared to baseline.
\item[\textbf{RQ2}] Users were able to enter queries 40\% faster compared to baseline.
\end{itemize}
\end{mdframed}

\section{Related Tools}

There have been a variety of approaches to designing and developing graphical query builders.

Film Finder~\cite{jog1995starfield} was an early example of using graphical user interface components to improve user accessibility. Film Finder created visual components for filtering each column of the dataset. \gqb~extends that research by creating a more complete query creation user interface. \gqb~can build more complex queries, such as multiple inclusion/exclusion ranges and areas, and works with spatial data. 

Two examples of a graphical query builders that work with spatial and frequency data are the Australian Communications and Media Authority (ACMA) and FM Query Broadcast Station Search by the Federal Communications Commission.\footnote{\url{https://www.fcc.gov/media/radio/fm-query}} Both these tools allow a user to search for registered transmitters within their respective countries. They both allow users to filter the data set through graphical components. The ACMA search also allows users to gradually build the query by adding search terms. Both tools require the user to submit the query before showing the results, and users do not have the option to continue adding conditions to an existing query, which are features supported by \gqb. These are both important tools in the area of spatio-temporal-frequency search, but have significant shortcoming that we felt could be improved on. 

Visual SQL, by Chartio,\footnote{\url{https://chartio.com/blog/why-we-made-sql-visual-and-how-we-finally-did-it/}} features a flexible drag-and-drop interface which allows users to explore and visualise data.  Users can import any dataset into the tool, which can automatically show the data structure to help users understand the data and enable them to construct queries. Once an initial query is built, users can quickly see the results in a chosen form (table, graph), modify the query, or narrow the output by adding conditions. While Visual SQL is a powerful tool, it is not customised for spatial, temporal, or spectral data, which is what motivated us to build \gqb. It would likely have been feasible to implement \gqb~ with Visual SQL, but we expect it would not have impacted the evaluation results.

Erwig and Schneider introduced STQL, a Spatio-Temporal Query Language~\cite{erwig2002stql}. However, the authors concede that a query language is not enough for end users because ``end users need an easy access to spatio-temporal data and queries''~\cite{erwig2002stql}. They point to visual querying as a potential solution. \gqb~builds on this research by creating a visual approach to query building. It includes a subset of SQTL functionality, just what was required by our industry partner. To the best of our knowledge STQL is not publicly available, which prevented us from using it as a baseline.

Many of these related tools show what can be done with graphical approaches to querying data. An important aspect of this paper is the measurement of the effectiveness of the approach. 

\section{Conclusions and Future Work}

\gqb~allows users to navigate through a large, complex dataset of spatial-temporal-spectral data with increased ease and simplicity. It is powerful to use, allowing users to query by transmitter name, location range, hours of operation, and frequency. Users can execute simple and complex queries more quickly, efficiently, and accurately than with the baseline tool. User can continually add queries based on their needs, and receive immediate visual feedback on their queries. We believe these features contributed to the results in improved user speed and accuracy, and positive user responses when querying the data. 

For future development of \gqb, our industry partner has expressed interest in:
\begin{itemize}
\item The ability to control `AND' and `OR' conditionals with brackets to allow more complex query structures.
\item A query interface that updates based on the imported data format. 
\item Import and export of more data formats.
\item More visual elements for constructing the queries.
\end{itemize}


\end{sloppy}
\end{document}